\documentstyle[floats,twocolumn,pra,aps]{revtex}
\input psfig.sty

\begin{document}


\title{Hydrodynamics of bacterial motion}
\author{Zolt\'an Csah\'ok and Andr\'as Czir\'ok}
\address{Department of Atomic Physics, E\"otv\"os University,
Puskin u. 5-7., H-1088 Budapest, Hungary.}
\date{\today}
\maketitle
\begin{abstract}
In this paper we  present a hydrodynamic approach to describe
the motion of migrating bacteria as a special class of self-propelled
systems.
Analytical and numerical calculations has been performed to
study the behavior of our model in the turbulent-like regime and to
show that a phase transition occurs as a function of noise strength.
Our results can explain previous experimental observations
as well as results of numerical simulations.
\end{abstract}

\bigskip
\pacs{Keywords: self-propelled, cooperative bacterial migration,
numerical simulation, phase transition.}

\section{Introduction}

Bacterial colony formation has been the subject of recent
studies performed by both
microbiologists\cite{Sha88,AH91,BB91,Sha95,BE95}
and physicists\cite{mat89,mat90,bj92,mat93,bj94,nature,BCSALT95,PRL2}
with the aim to
acquire a deeper understanding of the collective behavior of
unicellular organisms. 

Since bacterial colonies can consist of up to $10^{10}$
microorganisms,
one can distinguish several length and time scales
when describing their development.
At the largest, macroscopic scale (above cca. $10^{-3}$~m) one is
concerned about the morphology of the whole colony, which can be often
described in terms of fractal geometry \cite{mat89,mat93,nature}.
At this length scale, the {\it physical}~  laws of diffusion
govern colony formation, and only certain details of the
underlying microscopic dynamics (such as
anisotropy \cite{anis}) can be amplified up to this scale by various
instabilities.  Thus these
large-scale features could have been successfully understood using
relatively simple models which neglect most of the microscopic
details. In contrast, when looking at the microscopic length
scale of individual bacteria (cca. $10^{-6}$ m), such global physical
constraints play a much smaller role and it is the {\it biology} which
provides us the knowledge of how the microorganisms move,
multiplicate and communicate. In this work we focus on the
intermediate
length and time scales which range from $10^{-5}$~m to $10^{-3}$~m and
from $1$~s to  $10^2$~s. In this regime the proliferation of the
bacteria can be neglected,
and the most striking observable
phenomena are connected to the motion of the organisms.

Bacterial {\it swimming} has been well understood in liquid cultures
\cite{Adler,Berg,CHT}, where
the interaction between cells is usually negligible and the
trajectory of each organism can be described as a biased random walk
towards the higher concentration of the chemoattractants, e.g.,
certain nutrients. In fact, the bacterial motion consists of
straight sections separated by short periods of random rotation
called tumbling, when the flagellas operate in a quite
uncorrelated manner. If an increase of the local
concentration of the chemoattractant is detected, bacteria delay
tumbling and continue to swim in the same direction. This behavior
yields
the above mentioned bias towards the chemoattractant. Recent
experimental observations \cite{BB91,nature,BE95} revealed that under
stress conditions even chemotactic communication occurs in
colonies: chemoattractants or chemorepellents are emitted by the
bacteria
themself, thus controlling the development of various self-organized
structures\cite{BCSALT95}.

However, numerous bacterial strains like {\it Proteus
mirabilis, Serratia marcenses} \cite{AH91} {\it Bacillus
circulans, Archangium violanceum, Chondromyces apiculatus,
Clostridium tetani} \cite{film} or {\it Bacillus
subtilis} \cite{bj94} are also able to migrate on surfaces to
ensure their survival under hostile environmental conditions.
In all of these examples, irrespective of the cell membrane
(Gram $+$ or $-$) and the type of locomotion ({\it swarming} or
{\it gliding}), a significantly {\it coordinated} motion can be
observed: the randomness, which is a characteristic feature of
the {\it swimming} of an individual  organism, is rather repressed.
In particular the {\it swarmer}
cells of the strain {\it Proteus} are very elongated
(up to 50$\mu$m) and move parallel to each other \cite{AH91}.
The strain {\it Bacillus circulans} obtained its name from the
very complex flow trajectories exhibited, which include rotating 
droplets or rings made up of thousands of bacteria \cite{film,czPRE}.
To some extent similar behavior can be observed  in each of the
above listed examples, which naturally raises the long standing
question as to whether this form of migration requires some external
(or
self-generated) chemotactic signal or whether simple local cell-cell
interactions are sufficient to explain such correlated behavior.

One of the purposes of this paper is to show that the answer to such
questions requires taking into account that the bacteria are
far-from equilibrium, self-propelled systems.
Recent studies have
revealed that such systems show interesting and unexpected features:
in various traffic models
spontaneous jams appear \cite{Nag93,Ker93} and there is a strong
dependence on
quenched disorder \cite{Csa94}.
Many biological system (such as swarming bacteria, schools of fishes,
birds, ants, etc.) can be described as special self-propelled systems
with local velocity-velocity interaction introduced by Vicsek {\it et
al.}
\cite{Novel} and also studied in a lattice gas model \cite{Csz95}.
Both of these models revealed transition from disordered to ordered
phase.
Although most of the models above incorporate some level of
discreteness, a continuum approach also has been applied to study the
traffic
flow problem \cite{Ker93} or to describe the geotactic motion of
bacteria
\cite{Chi81}.
Toner and Tu  have shown using renormalization group theory
that in a closely related continuum model \cite{Tu95}
a long range order can develop even in two dimensions; this is
not possible in the equivalent equilibrium models. 

In this paper we construct a
continuum description for the particularly
interesting collective motion
in bacterial colonies.
First we introduce our model, then we give a theoretical analysis of
the problem. In Section\ \ref{sec_numres} we present the numerical
results
and finally in Section\ \ref{sec_concl} we draw our conclusions.

\section {The Model}

When migrating on surfaces, bacteria usually form a very thin film
consisting only of a single or a few layers of cells, thus their
motion can be regarded as quasi two-dimensional. Due to the
small length scale, the very small Reynolds number yields an
overdamped
dynamics in which the thermal fluctuations (Brownian noise)
and the various surface effects (e.g., surface tension, wetting)
dominate the motion. 
The constrains of the cell-cell interaction must also
be taken into account since  the elongated migrating bacteria usually
align
parallel to each other, which results in an
{\it effective viscosity} that tends to homogenize the velocities as
we
show later.
Bacteria also tend to move close to the substrate which
manifests in an {\it effective gravity}, flattening the colony.

\begin{figure}
\centerline{\psfig{figure=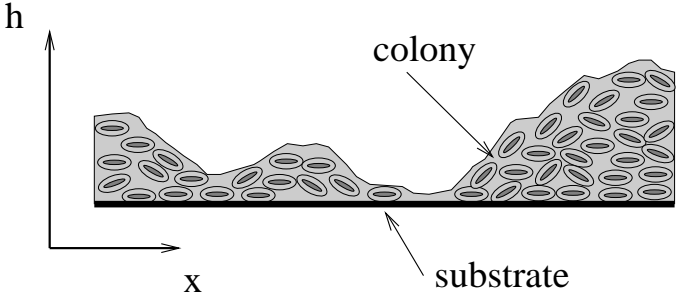,width=9cm}}
\caption{Schematic crossection of a bacterial colony.}
\label{bacifluid}
\end{figure}

The particles of the ``bacterial fluid'' (i.e., the colony)
are bacteria surrounded by a droplet of moisture
(extracellular fluid, see Fig.\ \ref{bacifluid}) which is essential
for them
both to move and  to
extract food from the substrate.
The volume ($V^*$) and the mass ($m^*$) of such a particle
is considered to be constant, which gives a constant (3D)
density $\varrho ^* = m^*/V^*$.
A typical bacterial colony grown on a surface can be considered as
flat,
so our fluid will be characterized by
the 2D {\it number} density $\varrho ({\bf r},t)$
and velocity ${\bf v} ({\bf r},t)$ fields, where ${\bf r}\in {\rm
R}^2$
and $t$ is the time.
The number density is simply related to the local height of the
colony $h$ by
$ h({\bf r},t) = \varrho ({\bf r},t) V^*$.

In contrast to fluid mechanics, here the fluid element consists of
only a few ``atoms''. Thus random fluctuations do not cancel
out and they have to be taken into account by an additional noise
term in the equation of motion for the velocity field.
The two basic equations governing the dynamics are
the continuity equation
\begin{equation}\label{cont0}
\partial_t \varrho + \nabla (\varrho {\bf v})= 0 
\end{equation}
and the equation of motion, which is a suitable form of the
Navier-Stokes
equation
\begin{eqnarray}\label{eom0} 
\partial_t {\bf v} + ({\bf v}\cdot \nabla) {\bf v} &=&
-{1\over \varrho^*} \nabla p + \nu \nabla^2 {\bf v} +
{1\over\varrho^*}{\bf F}({\bf v}) - {1\over\tau} {\bf v}  \nonumber\\
&&+{\bf \eta}({\bf x},t),
\end{eqnarray}
where $p$ is the pressure, $\nu$ is the kinematic viscosity,
${\bf F}({\bf v})$ is the {\it intrinsic} driving force of biological
origin,
$\tau$ is the time scale associated with substrate friction
and ${\bf \eta}({\bf x},t)$ is an uncorrelated fluctuating force
with zero mean and variance $\sigma$ representing the random
fluctuations.
Let us now go through the terms of Eq.\ \ref{eom0}.

The pressure is composed of the effective hydrostatic pressure,
the capillary pressure, and the externally applied pressure
\begin{equation}\label{press}
p = \varrho^* g h + \gamma \nabla^2 h + p_{\rm ext}.
\end{equation}
If the radius of surface curvature is larger
then the capillary length
\begin{equation}\label{lcap}
l_{\rm cap} = \sqrt{\gamma\over {\varrho^* g}}
\end{equation}
then the capillary pressure can be neglected.
Since this is normally  the case ($l_{\rm cap}\approx 3$ mm for
water),
we consider
only the first term of Eq.\ \ref{press} in the rest of the paper.
The surface tension becomes relevant only at the boundary of
the colony where the local curvature is not negligible.

The viscous term in Eq.\ \ref{eom0} represents not only the viscosity
of
the extracellular fluid, but also incorporates the local ordering of
the cells. As we show in Appendix A for a simple hard rod model of the
cell-cell interaction, the deterministic part of the dynamics of
the orientational ordering is described in a similar manner
to \cite{Novel}:
\begin{equation}
{{\rm d}\theta_i\over{\rm d}
t}=\mu\Bigl(\langle\theta\rangle_\epsilon-\theta_i\Bigr),
\label{orient}
\end{equation}
where the orientation of the $i$th rod is denoted by $0<\theta_i<\pi$
and $\langle\cdot\rangle_\epsilon$ denotes spatial averaging over a
ball 	
of radius $\epsilon$. 
The angle $\theta$ can be readily replaced by ${\bf e}_\theta$,
a unit vector at angle $\theta$ to the $x$ axis.
If the changes in the magnitude of the velocity are small
then ${\bf e}_\theta$ can be replaced by ${{\bf v}}$ and 
Eq.\ \ref{orient} yields an interaction term proportional to
$\langle {{\bf v}} \rangle_{\epsilon} - {{\bf v}}$.

Taking Taylor series expansions for the velocity
and the density fields yields 
\begin{eqnarray}
\langle {{\bf v}} \rangle_{\epsilon} - {\bf v} &=&
{
\int_{\vert {\bf \xi} \vert < \epsilon} \hbox{d} {\bf \xi} \Bigl(
{{\bf v}}\varrho+
({\bf \xi}\nabla) {{\bf v}} \varrho +{1\over 2}({\bf \xi}\nabla)^2
{{\bf v}}\varrho
+ \dots \Bigr)
\over
\int_{\vert {\bf \xi} \vert < \epsilon} \hbox{d} {\bf \xi} \Bigl(
\varrho + ({\bf \xi}\nabla)\varrho +{1\over 2}({\bf
\xi}\nabla)^2\varrho
+ \dots \Bigr) } - {\bf v} \nonumber\cr 
& =&  {\epsilon^2\over 6}\Bigl(
{\nabla^2 ({{\bf v}}\varrho)\over \varrho} -
{{\bf v}}{\nabla^2 \varrho \over \varrho} \Bigr) + \dots \nonumber\cr
&=&
{\epsilon^2\over 6}\Bigl( \nabla^2{{\bf v}} +
2(\nabla {{\bf v}}) {\nabla\varrho\over\varrho} \Bigr) + \dots
\label{DIFF1}
\end{eqnarray}
If the density changes are small, we recover the viscous term of Eq.\
\ref{eom0}, 
so it, in fact, includes the self-alignment rule of previous models.

Bacteria tend to maintain their motion continuously
by propelling with their flagellas. This rather complex behavior can
be taken
into account as a constant magnitude force acting
in the direction of their velocity
\begin{equation}
{\bf F} = \varrho^* {c\over\tau}{{{\bf v}}\over{\vert {{\bf v}}
\vert}},
\label{sdrive}
\end{equation}
where $c$ is the speed determined by the balance of the propulsion
and friction forces. That is, $c$ would be the speed of a homogeneous
fluid.  

Combining  Eqs.\ \ref{cont0}, \ref{eom0}, \ref{press} and
\ref{sdrive}  we obtain the following final form for the equations
of the bacterial flow:
\begin{equation}
\label{cont}
\partial_t h + \nabla (h {\bf v})= 0, 		
\end{equation}
and
\begin{eqnarray}
\label{eom}
\partial_t {\bf v} + ({\bf v}\cdot \nabla) {\bf v} &=& -{g} \nabla h +
-{1\over\varrho^*}\nabla p_{\rm ext} +
\nu \nabla^2 {\bf v} \nonumber\\
&&+{c\over\tau}{{{\bf v}}\over{\vert {{\bf v}} \vert}} - {1\over\tau}
{\bf v} + {\bf \eta}.
\end{eqnarray}
These equations are similar to those studied in \cite{Tu95} but 
here we derived them from plausible assumptions based on the
underlying 
microscopic dynamics instead of phenomenological concepts. 
Nevertheless, the analysis of Toner and Tu also holds for our
equations 
which permits us to use our model as a test of their prediction of 
the existence of a phase transition. 

\section{Analytical results}

For certain simple geometries of the boundary condition
it is possible to obtain analytical solutions for the noiseless
($\sigma=0$) stationary state
of our model if we suppose incompressibility ($h=const.$).
Taking $\partial_t \equiv 0$ in Eqs.\ \ref{cont} and \ref{eom},
the following dimensionless equations are obtained:
\begin{equation}
\nabla' {{\bf v}}' = 0
\label{cont2}
\end{equation}
and
\begin{equation}
{{c \tau}\over{\lambda}}
({\bf v}' \nabla') {\bf v}' =
-{1\over{\varrho^* \lambda}}\nabla p_{\rm ext} + {\nabla}'^2 {\bf v}'
+
{{\bf v}'\over{\vert {{\bf v}}' \vert}} -  {{{\bf v}}}',
\label{eom2}
\end{equation}
where ${{\bf v}}'={{\bf v}}/c$, $\lambda=\sqrt{\nu\tau}$
and the $\nabla'$ operator derivates
with respect to ${\bf r}'={\bf r}/\lambda$.
For the sake of simplicity we drop the prime in the rest of this
section. 

First let us consider the simplest geometry when the system is defined
on an infinite plane. In this case the stationary state is
trivially
\begin{equation}
\vert {{\bf v}}({\bf r}) \vert = 1,
\end{equation}
where the orientation of ${{\bf v}}$ is arbitrary, but independent of
the spatial position.
In this state there is a sustained non-zero
net flux of the fluid so it is regarded to be {\it ordered}.
In the next section we examine numerically how
increasing the noise level yields a disordered state where there is
no net flux present.

Another simple, but practically more relevant boundary condition is
realized when the system is confined to a circular area of radius $R$.
In contrary to the previous example this is a finite geometry
which in turn means that in the stationary
state no net flux is possible. Thus the flow field must include
vortices and we  
show that a single vortex is indeed
a possible stationary configuration of the system. 

Let us assume that the velocity field,
expressed in polar coordinates $(r,\phi)$,
is a function only of $r$:
\begin{equation}
{{\bf v}} = v(r)~{\bf e}_\phi,
\label{vpdef}
\end{equation}
where ${\bf e}_\phi$ is the tangential unit vector.
Taking the expression of the nabla operator in polar coordinates,
one can easily see that Eq.\ \ref{cont2} is satisfied for any
velocity profile $v(r)$.
Substituting Eq.\ \ref{vpdef} into Eq.\ \ref{eom2} yields two
ordinary differential
equations:
\begin{equation}
0= r^2 {{{\rm d}^2 v}\over {\rm d} r^2} + r {{{\rm d} v}\over {\rm d}
r}
- v (1+r^2) + r^2,
\end{equation}
and
\begin{equation}
-{1\over r^2}v^2 =
-{1\over{\varrho^* \lambda}}{{\partial p_{\rm ext}}\over \partial r}.
\end{equation}
The first equation gives the velocity profile while the second
determines the pressure to be applied for maintaining the constant
height
($h=$const.) condition.

\begin{figure}
\centerline{\psfig{figure=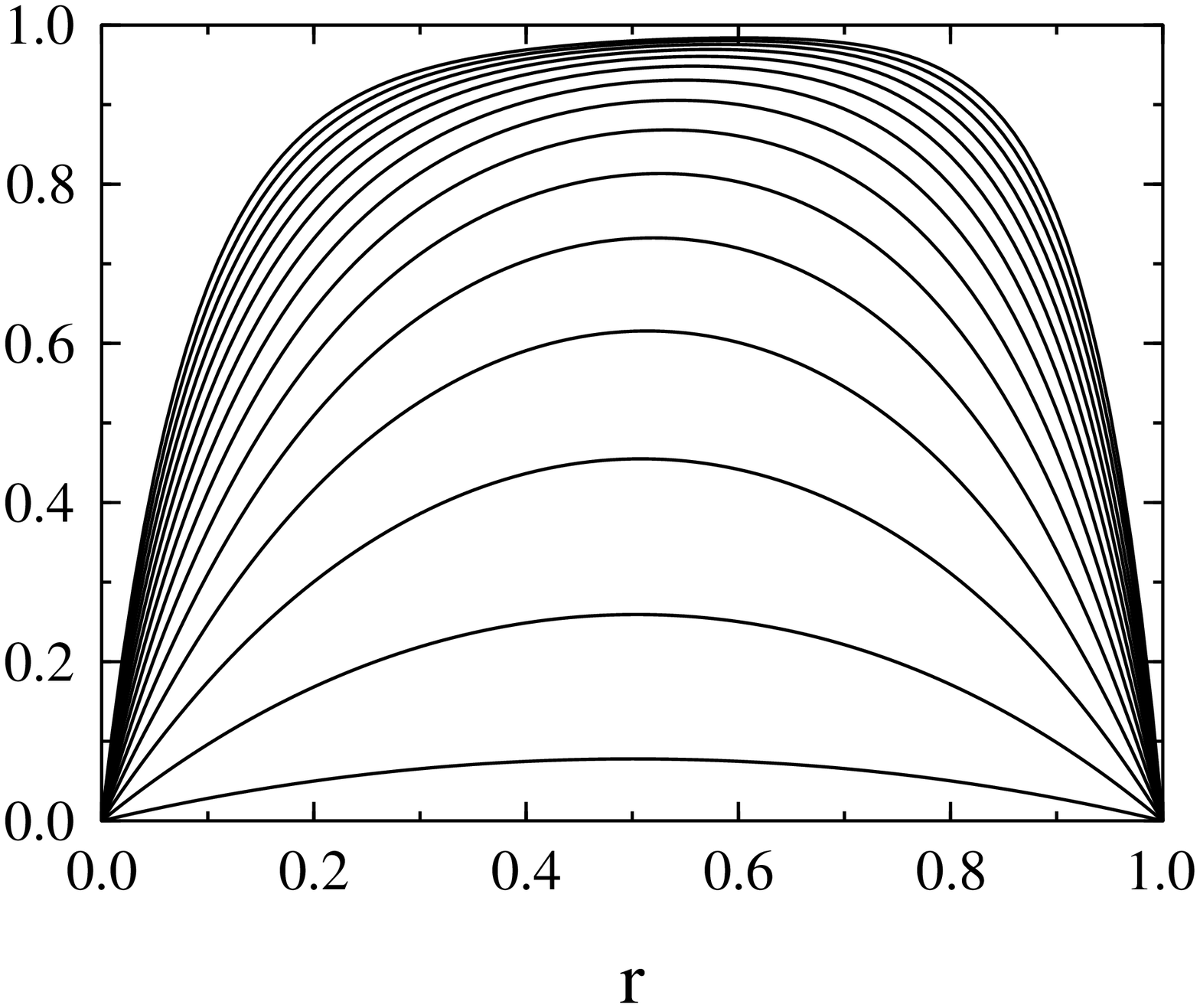,width=9cm}}
\vspace{0.5cm}
\caption{Velocity profiles in a vortex
for various values of ${\rm Re}'$ (bottom curve: ${\rm Re}'=1$,
top curve: ${\rm Re}'=16$).}
\label{velprofs}
\end{figure}

\begin{figure}
\centerline{\psfig{figure=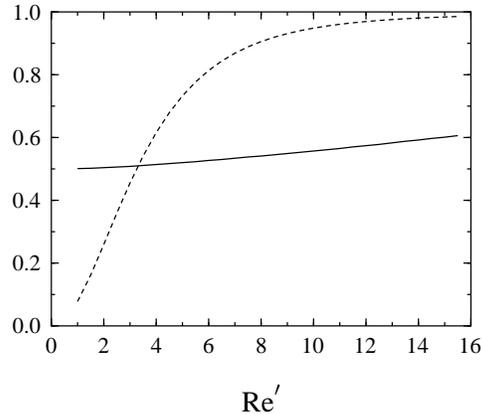,width=8cm}} 
\vspace{0.5cm}
\caption{Position (solid line) and value (dashed line)
of the maximal velocity for the profiles on
\protect Fig.\ \ref{velprofs}
as a function of ${\rm Re}'$. At the maximal velocity, open boundary
condition
(${\rm d} v /{\rm d} r =0$) is satisfied.}
\label{vmax}
\end{figure}

The boundary conditions for the velocity profile are either
$v(0)=0$ and $v(R')=0$ (closed boundary at $r=R'=R/\lambda$)
or $v(0)=0$ and ${{{\rm d} v}\over{{\rm d} r}}(R')=0$ (free boundary
at $r=R'$).
The homogeneous solution of the equation of motion with
the above boundary conditions is
given by $I_1(r)$, the modified Bessel function of order one.
The particular solution is
$-{\pi\over 2}L_1(r)$, where $L_1$ is the modified Struve function of
order one \cite{Stegun}.
Thus the velocity profile of the single vortex
stationary state in a noiseless system with circular boundary is
\begin{equation}
v(r) = \alpha I_1(r) -{\pi\over 2}L_1(r).
\label{vp}
\end{equation}
The parameter $\alpha$ should be chosen to satisfy
the boundary condition at $r=R'$.
In Fig.\ \ref{velprofs} we show velocity profiles
for different values of $\lambda$ having $R=1$ fixed and
imposing closed boundary condition.
The maximal velocity decreases for decreasing $R/\lambda$ ratio
(Fig.\ \ref{vmax}); thus the system behaves more similar to the usual
(not self-propelled) systems where $v(r)\equiv 0$.
In this respect $R/\lambda$ is a measure of
the self-propulsion. From Fig.\ \ref{velprofs} it is also clear
that the minimal size
of a vortex is $\lambda$ so if $R/\lambda\gg 1$ then
many vortices are likely to be present in the system.
This phenomenon is analogous to the appearance of turbulent
motion in fluid dynamics and thus $R/\lambda$ can be regarded as a
quantity
analogous to the Reynolds number:
\begin{equation}
{\rm Re}' = {R\over\lambda}.
\end{equation}
Another dimensionless quantity characterizing the flow
is $c\tau/\lambda$ which gives the relative strength of
the inertial forces compared the viscous ones. This quantity
can be regarded as an internal Reynolds number
\begin{equation}
{\rm Re}'' = {c\tau\over\lambda}.
\end{equation}
In our
simulations we always had ${\rm Re}'' \ll {\rm Re}'$.

\section{Numerical results}
\label{sec_numres}

For further investigations we used numerical solutions of Eq.\
\ref{cont}
and Eq.\ \ref{eom}.
To study the closed circular geometry we implemented our model
on a hexagonal region of a triangular lattice with closed boundary
conditions. We used a simple explicit integration scheme to solve
numerically
the equations.

\begin{figure}
\centerline{\psfig{figure=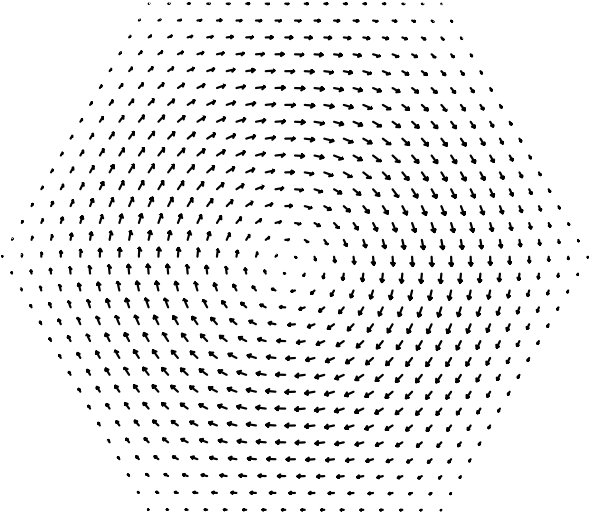,width=8cm}}
\caption{A numerically generated vortex for
${\rm Re}'\simeq 4.4$ and $g= 750$. The length of the arrows is
proportional
to the local velocity, while their thickness is proportional to the
density.}
\label{numvort}
\end{figure}

\begin{figure}
\centerline{\psfig{figure=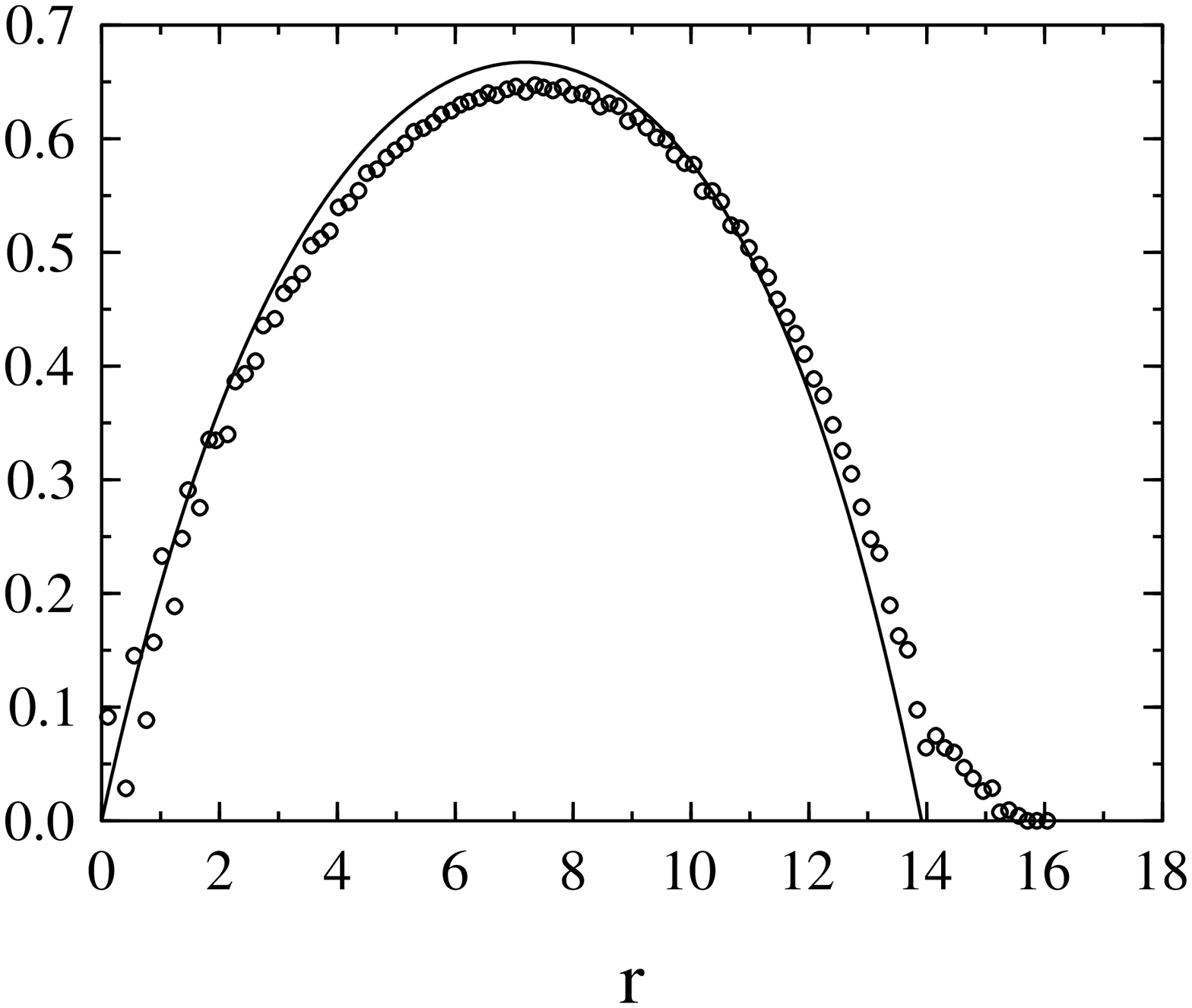,width=8cm}}
\vspace{0.5cm}
\caption{The measured (circles) and
the theoretical (solid line)
velocity profile for the vortex in \protect Fig.\ \ref{numvort}.}
\label{numfit}
\end{figure}

\begin{figure}
\centerline{\psfig{figure=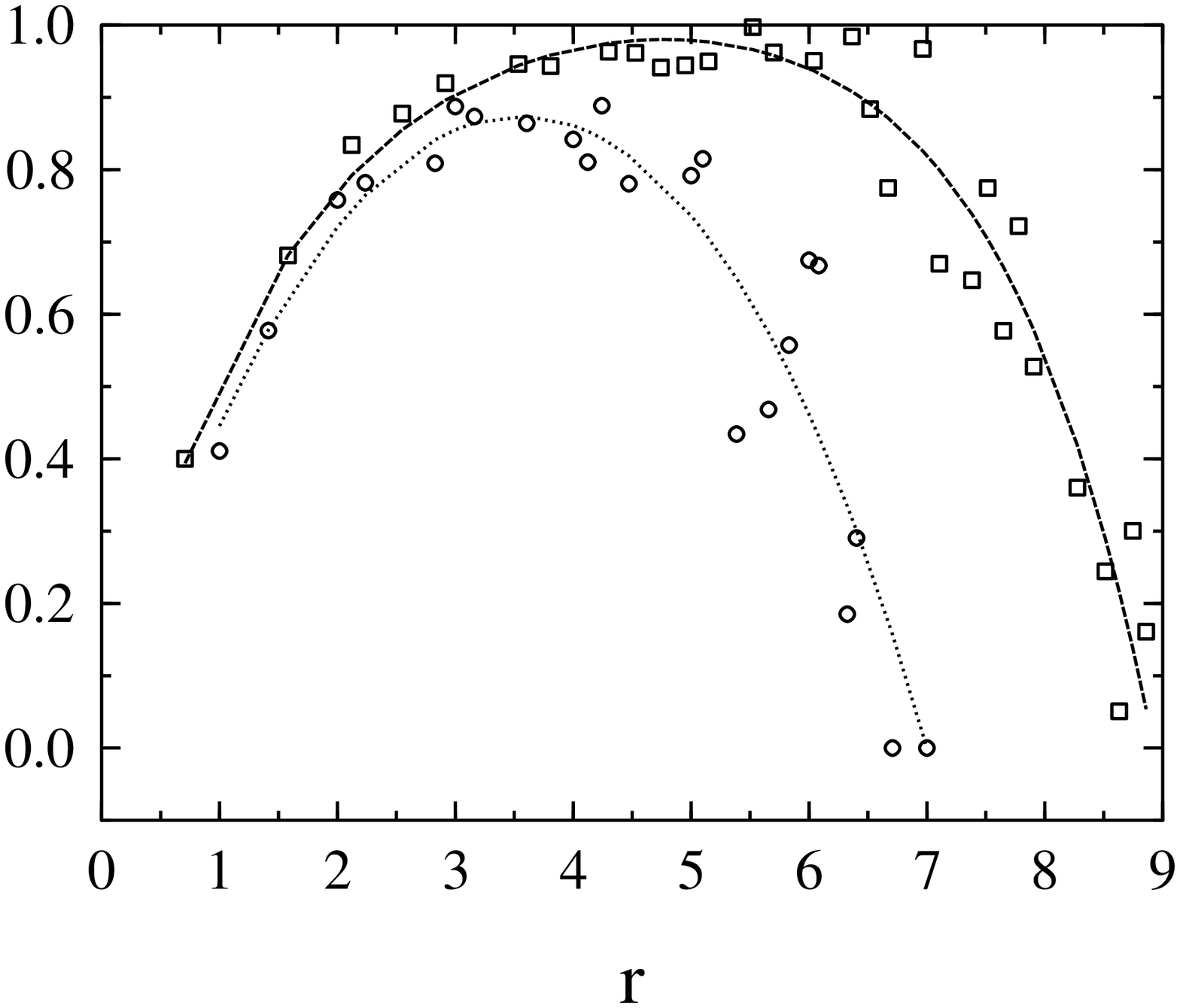,width=8cm}}
\vspace{0.5cm}
\caption{Velocity profile data for two vortices taken from
\protect \cite{czPRE} (circles and squares) and
fitted profiles using \protect Eq.\ \ref{vp}.}
\label{bacifit}
\end{figure}

We started our simulations
from a uniform density and a random velocity distribution.
Figure\ \ref{numvort} shows the stationary state for the high
viscosity
($\lambda\simeq 3.16$)
and high compressibility ($g=750$) case, corresponding to
${\rm Re}'\simeq 4.4$.
The length and direction of the arrows show the velocity, while
the thickness is proportional to the local density of the fluid.
In Fig.\ \ref{numfit} we present the radial velocity distribution for
the vortex
shown in Fig.\ \ref{numvort} and the velocity profile given by our
calculations
(Eq.\ \ref{vp}). Rather good agreement is seen; the differences are
due to the
fact that our numerical system is not perfectly circular.
We have also performed velocity profile analysis for two vortices from
\cite{czPRE}.
The data and the fitted curves are displayed in Fig.\ \ref{bacifit}.
Again, reasonable agreement is seen which  means that Eqs.\ \ref{cont}
and \ref{eom} give a correct continuous description for this
particular example of self-propelled systems.

\begin{figure}
\centerline{\psfig{figure=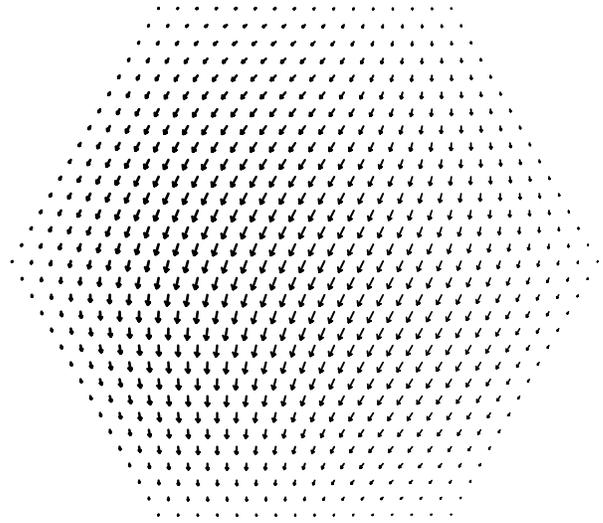,width=8cm}}
\caption{Non-stationary configuration at ${\rm Re}'= 4.4 , g= 300$.}
\label{wash}
\end{figure}

In the case of real bacteria one hardly can observe
a perfect vortex, so we tuned the parameters to see
the behavior of the model far from stationarity.
Lowering the value of the compressibility $g$, we observed temporally
periodic structures instead of a constant velocity profile vortex.
Figure\ \ref{wash} shows such a configuration.
The density at the lower
left part of the system is higher than at the top right part.
As the system evolves in time the whole flow pattern
rotates anti-clockwise, thus
this state has non-zero net flux which oscillates in time. 
Similar behavior also has been reported for certain  
bacterial colonies\cite{bj94} . 

\begin{figure}
\centerline{\psfig{figure=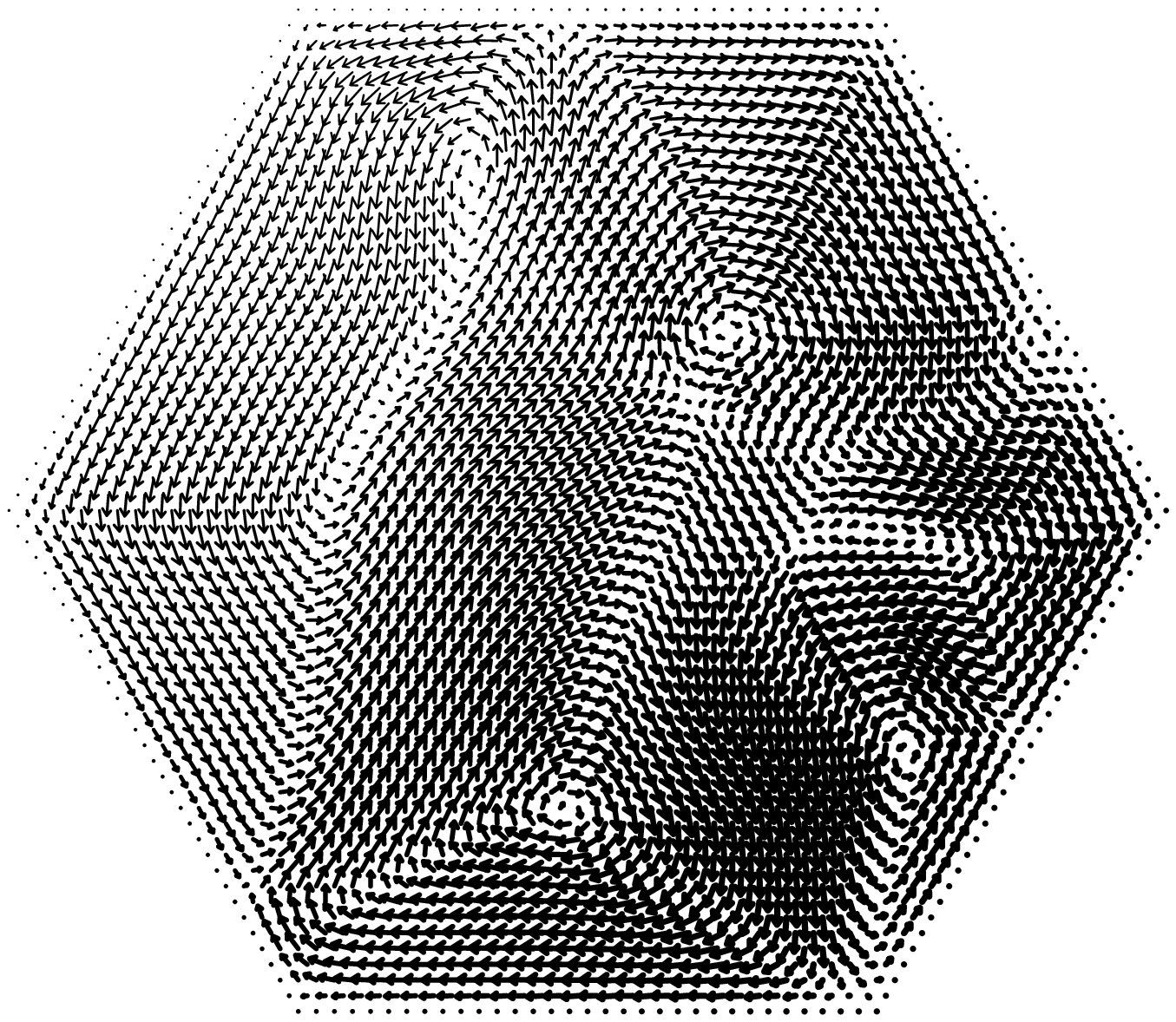,width=8cm}}
\caption{Multiple vortex state at ${\rm Re}'= 25.3 , g= 128$.}
\label{xxvortex}
\end{figure}

Another interesting case occurs when the compressibility is high and
the viscosity is low, which corresponds to high values of ${\rm Re}'$.
In this limit we observed a long lifetime multi-vortex state
(Fig.\ \ref{xxvortex}). For the parameter values we use, these
vortices
eventually disappeared leaving behind a single vortex. We conjecture
that there exists a ${\rm Re}'_c$ above which the multi-vortex
state does not die out, this would correspond to the turbulent
state of the normal hydrodynamics.

\begin{figure}
\centerline{\psfig{figure=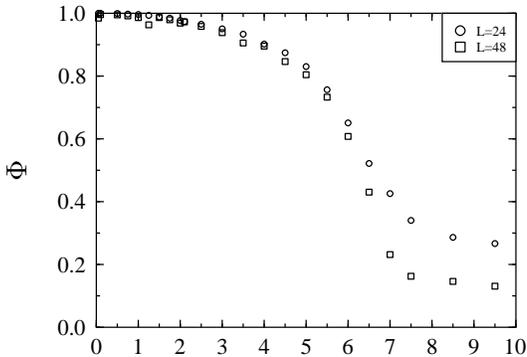,width=8cm}}
\caption{The order parameter $\Phi$ as a function of
the noise strength $\sigma$.}
\label{phtrans}
\end{figure}

We also performed simulations in the open planar geometry.
In this case our numerical
system was confined to a square with periodic boundary conditions. 
Our goal was to find if there exists the phase transition
as a function of the noise strength $\sigma$.
To do this we first
defined an order parameter
\begin{equation}
\Phi = \langle \vert {\bf v} \vert \rangle,
\end{equation}
where $\langle\cdot\rangle$ denotes spatial average over the entire
system.
If $\Phi=0$ there is no net current and the system 
is in disordered state. If $\Phi>0$ some level
of order is present.
We have performed long time ($>10^7$ Monte-Carlo steps) 
runs for two different system sizes
(L=24 and 48) to check
the presence of the transition. Figure\ \ref{phtrans} shows the
results obtained
which strongly suggests that there is a transition around
$\sigma\approx 6.5$. This value is by far not universal as it depends
on 
the actual values of parameters used in the simulations.
However, the extraction of the critical exponents  would require much
larger
computational efforts (i.e., larger system sizes).

\section{Conclusion}
\label{sec_concl}

We have investigated a continuous model for cooperative
motion of bacteria. For simple geometries
and specific parameters we were able to obtain analytical
results for the velocity profile of vortices often observable
on intermediate length scales in various bacterial colonies.
Our results are in quantitative agreement with biological
observations \cite{film} and with previous simulations
\cite{czPRE}. We also showed that there exists a transition
\cite{Tu95}
from an ordered to a disordered phase as a function of noise
strength.

In its present form our model does not give a model for the
pattern formation of bacterial colonies. However, it is rather
straightforward to include surface tension and wetting effects in
their
full detail \cite{Gen85} to handling the dynamics of the colony
border. Including proliferation, the presented model can be also
expanded towards larger time and length scales to study the fingering
instabilities. It is also possible to introduce chemotactic response
which would allow long range interactions to build up.

Further work needs to be done to characterize the turbulent regime
(which is mostly observed in swarming colonies)
and to relate the effective parameters used in our model
to the real world quantities such as food or agar  concentration.  

\section{Acknowledgments}
The authors thank E. Ben-Jacob, H. Herrmann and  T. Vicsek for useful
discussions and comments.
This work was supported by contracts T019299 and T4439
of the Hungarian Science Foundation
(OTKA) and by the ``A Magyar Tudom\'any\'ert'' foundation of
the Hungarian Credit Bank.

\newpage

\appendix
\section{Simple geometrical model for the orientational ordering of
cells}

In this appendix we show that the simple geometrical constrains
originated from the rod-like shape of bacteria can motivate the
local orientational ordering introduced in Sec. II.
Microscopic observations of high density cell cultures reveal
that the orientational order emerges since bacteria tend not to
``intersect'' each other: usually bacteria form
well defined layers in which they are parallel aligned. To
investigate quantitatively this phenomenon let us consider a
system of rods with unit length and negligible width, covering
the plane with a uniform average density $\varrho$. Here we
focus on the local orientational order only and neglect the
translational movement of the cells. Thus let us assume that
the only interaction in the system is the intersection of the
rods, and the overdamped dynamics of the system is determined by
two effects, a relaxation process minimizing the total number of
intersections $U$ and a random Brownian fluctuation as
\begin{equation}
{{\rm d}\theta_i\over{\rm d} t} =
-\lambda{{\partial}\over{\partial}\theta_i}U(\theta_1,\theta_2,\dots)
+ \xi_i \hbox{~~mod~~} \pi,
\label{A1}
\end{equation}
where the direction of the $i$-th rod is denoted by $\theta_i$ and
$\xi_i$ is an uncorrelated noise: $\langle\xi_i(t)\xi_j(t')\rangle =
C\delta_{ij}\delta(t-t')$. As the rods are not directed,
$0<\theta_i<\pi$. The quantity $U$ can be written as a sum of ``pair
potentials'' $V_{ij}$:
\begin{equation}
U(\theta_1,\theta_2,..)={1\over 2}\sum_{i\not=j}
V_{ij}({{\bf x}}_i,\theta_i,{{\bf x}}_j,\theta_j),
\end{equation}
where $V_{ij}$ equals $1$ if the rods $i$ and $j$ intersect
and $0$ otherwise.

\begin{figure}
\centerline{\psfig{figure=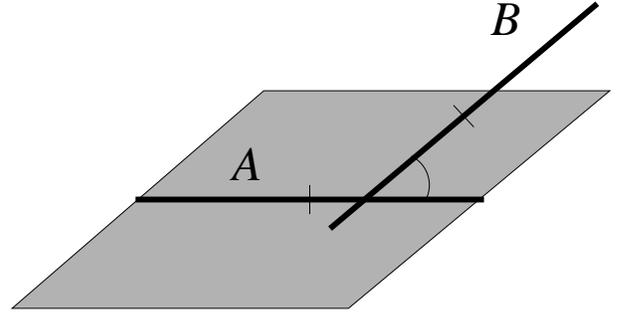,angle=-90,width=8cm}}
\vspace{0.5cm}
\caption{The rods $A$ and $B$ with a given orientation intersect
each other if the center of rod $B$ is inside of the displayed
parallelogram attached to rod $A$.
}
\label{andrasfig}
\end{figure}

As a mean-field approximation, we introduce the probability
density function $P(\theta)$, which gives the probability density
of the event that the direction of a randomly selected rod is in
the interval $[\theta,\theta+{\rm d}\theta]$. We are interested
in the local (short range) ordering; thus the spatial homogeneity of
$P$ is assumed.  Now $U(\theta)$, the expected number of intersections
of a rod directed at angle $\theta$, can be expressed as
\begin{equation}
U(\theta)=\varrho \int_V {\rm d}^2{{\bf x}}' \int_0^\pi {\rm d}\theta'
V({{\bf x}}_i,\theta,{{\bf x}}',\theta') P(\theta'),
\label{A2}
\end{equation}
since $\varrho P(\theta)$ gives the probability density for the
existence of a rod pointing in the direction $\theta$ at any position.
Due to the translational and rotational invariance of the system,
$U(\theta)$ is given in the form of
\begin{equation}
U(\theta)=\varrho \int_0^\pi {\rm d}\theta' P(\theta') \int_V {\rm
d}^2{{\bf x}}'
V(0,0,{{\bf x}}',\theta'-\theta).
\label{A3}
\end{equation}
As $\int {\rm d}^2{{\bf x}}' V(0,0,{{\bf x}}',\theta'-\theta)$ gives
the area of a
parallelogram shown in Fig.\ \ref{andrasfig}, (\ref{A3}) can be
written as
\begin{equation}
U(\theta)=\varrho \int_0^\pi {\rm d}\theta' P(\theta') \vert
\sin(\theta'-\theta) \vert.
\label{A4}
\end{equation}

The time evolution of the probability density function is given
by the Fokker-Planck equation obtained from (\ref{A1}):
\begin{equation}
{{\partial} P\over{\partial} t}={{\partial} \over{\partial}\theta}
\Bigl(\lambda P
{{\partial}\over{\partial}\theta} U +
{C\over2}{{\partial}\over{\partial}\theta} P \Bigr).
\label{FP}
\end{equation}
The steady state solution of Eq.\ \ref{FP}
assuming detailed balance satisfies
\begin{equation}
\lambda P{{\partial} U\over{\partial}\theta} + {C\over2} {{\partial}
P \over{\partial}\theta} = 0,
\end{equation}
yielding
\begin{equation}
P(\theta) = P_0\exp\Bigl(-{2\lambda U(\theta)\over C}\Bigr),
\label{SS}
\end{equation}
where the normalization for $P(\theta)$ is provided by the
prefactor $P_0$.

Let us measure $\theta$ compared to the spontaneously selected
orientation $\langle \theta \rangle$. For small deviations
$U(\theta)$ can be expanded in a Taylor series as
\begin{equation}
U(\theta) \approx {A\over2} \theta^2 + O(\theta^4).
\label{taylor}
\end{equation}
Now, from Eqs.\ \ref{SS} and \ref{A4} the coefficient $A$
can be calculated for given $\varrho$, $\lambda$ and $D$.
Substituting Eq.\ \ref{taylor} into Eq.\ \ref{SS} gives
\begin{equation}
P(\theta) \approx P_0 e^{-A\lambda\theta^2/C},
\end{equation}
with
\begin{equation}
P_0=\sqrt{A\lambda\over\pi C}.
\end{equation}
For $\theta, \theta'\ll 1$ the $\vert\sin(\theta'-\theta)\vert$ term
of
Eq.\ \ref{A4} can be linearized both in $\theta$ and $\theta'$:
\begin{eqnarray}
U(\theta)&=&
\varrho\theta\left( \int_{-\pi/2}^{\theta} d\theta' P(\theta') -
\int_{\theta}^{\pi/2} d\theta' P(\theta') \right)\nonumber\\
&& - \left(
\int_{-\pi/2}^{\theta} d\theta' P(\theta') \theta' -
\int_{\theta}^{\pi/2} d\theta' P(\theta') \theta' \right).
\end{eqnarray}
After some algebraic manipulations the above expression can be
simplified to
\begin{equation}
U(\theta)=2\varrho\int_{0}^\theta{\rm
d}\theta'P(\theta')(\theta-\theta')
\approx \varrho P_0\theta^2
\end{equation}
yielding the consistency condition $\varrho P_0=A/2$. Thus, in
this mean field approximation
\begin{equation}
{{\rm d}\theta_i\over{\rm d} t}= - {\lambda\over\pi C}
(\theta_i-\langle\theta\rangle_i),
\end{equation}
that is we recover Eq.\ \ref{orient}.
The angle $\theta$ can be readily replaced by ${\bf e}_\theta$,
a unit vector at angle $\theta$ to the $x$ axis.
If the changes in the magnitude of the velocity are small
then ${\bf e}_\theta$ can be replaced by ${{\bf v}}$ and 
we recover the self-aligning interaction Eq.\ \ref{orient}.

\end{document}